\documentclass[runningheads]{llncs}
\usepackage[pdftex]{graphicx,xcolor}
\usepackage[fleqn]{amsmath}
\usepackage{amsmath}
\usepackage{newtxtext}
\usepackage[varg]{newtxmath}
\usepackage{latexsym}
\usepackage{epsfig}
\usepackage{amssymb}
\usepackage{listings}
\usepackage{float}
\usepackage[framemethod=tikz]{mdframed}
\usepackage{environ}
\usepackage{tabularx}
\usepackage{algorithm}
\usepackage[noend]{algpseudocode}
\usepackage{syntax}
\usepackage{multirow}
\usepackage{tikz}
\usepackage{url}

\makeatletter

\newcommand{\problemtitle}[1]{\gdef\@problemtitle{#1}}
\newcommand{\probleminput}[1]{\gdef\@probleminput{#1}}
\newcommand{\problemquestion}[1]{\gdef\@problemquestion{#1}}

\makeatletter
\def\BState{\State\hskip-\ALG@thistlm}
\makeatother

\algnewcommand\algorithmicforeach{\textbf{for each}}
\algdef{S}[FOR]{ForEach}[1]{\algorithmicforeach\ #1\ \algorithmicdo}

\algnewcommand\algorithmicswitch{\textbf{switch}}
\algnewcommand\algorithmiccase{\textbf{case}}
\algnewcommand\algorithmicassert{\texttt{assert}}
\algnewcommand\Assert[1]{\State \algorithmicassert(#1)}%
\algdef{SE}[SWITCH]{Switch}{EndSwitch}[1]{\algorithmicswitch\ #1\ \algorithmicdo}{\algorithmicend\ \algorithmicswitch}%
\algdef{SE}[CASE]{Case}{EndCase}[1]{\algorithmiccase\ #1}{\algorithmicend\ \algorithmiccase}%
\algtext*{EndSwitch}%
\algtext*{EndCase}%
\let\oldReturn\Return
\renewcommand{\Return}{\State\oldReturn}

\newcommand{\Bool}{\mathbb{B}}
\newcommand{\True}{\top}
\newcommand{\False}{\bot}
\newcommand{\Nat}{\mathbb{N}}
\newcommand{\Natz}{\mathbb{N}_0}
\newcommand{\Rat}{\mathbb{Q}}
\newcommand{\pre}{\mbox{pre}}
\renewcommand{\epsilon}{\varepsilon}
\newcommand{\ptrans}[1]{\stackrel{#1}{\to}}
\newcommand{\QIF}{\mbox{QIF}}
\newcommand{\QIFdyn}{\mbox{QIF}^{dyn}}
\newcommand{\QIFone}{\mbox{QIF1}}
\newcommand{\QIFtwo}{\mbox{QIF2}}
\newcommand{\CQIFone}{\mbox{CompQIF1}}
\newcommand{\CQIFtwo}{\mbox{CompQIF2}}
\newcommand{\BEL}{\mbox{QIF}^{belief}}
\begin{document}
%
\title{Quantifying Dynamic Leakage {\it \\ \large Complexity Analysis and Model Counting-based Calculation }}
%
%
\author{Bao Trung Chu \and
Kenji Hashimoto \and
Hiroyuki Seki}
\authorrunning{B.\ T.\ Chu et al.}
\titlerunning{Quantifying Dynamic Leakage}
%
\institute{Graduate School of Information Science, Nagoya University, Japan}
\maketitle              
\begin{abstract}

	A program is non-interferent if it leaks no secret information to an observable output. 
However, non-interference is too strict in many practical cases and 
quantitative information flow (QIF) has been proposed and studied in depth. 
Originally, QIF is defined as the average of leakage amount of secret information
over all executions of a program. 
However, a vulnerable program that has executions leaking the whole secret 
but has the small average leakage could be considered as secure. 
This counter-intuition raises a need for 
a new definition of information leakage of a particular run, \textit{i.e., dynamic leakage}. 
As discussed in \cite{Bi16}, entropy-based definitions do not work well 
for quantifying information leakage dynamically; 
Belief-based definition on the other hand is appropriate for deterministic programs, 
however, it is not appropriate for probabilistic ones. 

In this paper, we propose new simple notions of dynamic leakage based on entropy
which are compatible with existing QIF definitions for deterministic programs, and yet
reasonable for probabilistic programs in the sense of \cite{Bi16}. 
We also investigated the complexity of computing the proposed dynamic leakage for three classes of Boolean programs.
We also implemented a tool for QIF calculation using model counting tools for Boolean formulae.
Experimental results on popular benchmarks of QIF research show the flexibility of our framework. Finally, we discuss the improvement of performance and scalability of the proposed method as well as an extension to more general cases.	

\keywords{Quantitative information flow  \and Hybrid monitor \and Dynamic leakage.}
\end{abstract}

\section{Introduction}
Researchers have realized the importance of knowing where confidential information reaches 
by the execution of a program to verify whether the program is safe. 
The {\it non-interference} property, namely, any change of confidential input does not affect public output, 
was coined in 1982 by Goguen and Meseguer \cite{GM82} as a criterion for the safety. 
This property, however, is too strict in many practical cases, 
such as password verification, voting protocol and averaging scores. 
A more elaborated notion called quantitative information flow (QIF) \cite{Sm09}
has been getting much attention of the community. 
QIF is defined as the amount of information leakage from secret input to observable output. 
The program can be considered to be safe (resp. vulnerable) if this quantity is negligible (resp. large). 
QIF analysis is not easier than verifying non-interference property 
because if we can calculate QIF of a program, 
we can decide whether it satisfies non-interference or not. 
QIF calculation is normally approached in an information-theoretic fashion 
to consider a program as a communication channel with input as source, and output as destination. 
The quantification is based on entropy notions
including Shannon entropy, min-entropy and guessing entropy \cite{Sm09}. 
QIF (or the information leakage) is defined as the remaining uncertainty about secret input 
after observing public output, i.e., the mutual information between source and destination of the channel. 
Another quantification proposed by Clarkson, et al. \cite{CMS09}, is the difference between `distances' (Kullback-Leibler divergence) 
from the probability distribution on secret input that an attacker believes in to the real distribution, 
before and after observing the output values.
\newline
While QIF is about the average amount of leaked information over all observable outputs, \textit{dynamic leakage} is about
the amount of information leaked by observing a \textit{particular} output. Hence, QIF is aimed to verify the safety of a program in
a static scenario in compile time, and dynamic leakage is aimed to verify the safety of a specific running of a program. 
So which of them should be used as
a metric to evaluate a system depends on in what scenario the software is being considered.

\lstset{
	basicstyle = \small\normalfont,
	escapechar=\%
}
\begin{example}
	\label{ex1.1}
\end{example}
\begin{quote}
	if $source<16$ then $output \gets 8 + source$
	\newline
	else $output \gets 8$
\end{quote}
In Example \ref{ex1.1} above, assume $source$ to be a positive integer, 
then there are 16 possible values of $output$, from 8 to 23. 
While an observable value between 9 and 23 reveals {\it everything} about the secret variable, 
i.e., there is only one possible value of $source$ to produce such $output$, 
a value of 8 gives almost nothing, 
i.e., there are so many possible values of $source$ which produce 8 as output. 
Taking the average of leakages on all possible execution paths results in a relatively small value, 
which misleads us into regarding that the vulnerability of this program is small. 
Therefore, it is crucial to differentiate risky execution paths from safe ones 
by calculating dynamic leakage, i.e., the amount of information can be learned from observing the output 
which is produced by a specific execution path.
But, as discussed in \cite{Bi16}, 
any of existing QIF models (either entropy based or belief tracking based) does not always seem reasonable 
to quantify dynamic leakage. 
For example, entropy-based measures give sometimes negative leakage. Usually, 
we consider that the larger the value of the measure is, the more information 
is leaked, and in particular, no information is leaked when the value is 0. 
In the interpretation, it is not clear how we should interpret a negative value 
as a leakage metric. Actually, \cite{Bi16} claims that the non-negativeness is a requirement for 
a measure of dynamic QIF. Also, \textit{MONO}, one of the axioms for QIF in \cite{ACM16} 
turns out to be identical to this non-negative requirement.
Belief-based one always give non-negative leakage for deterministic programs
but it may become negative for probabilistic programs. 
In addition, the measure using belief model depends on secret values. This would imply
(1) even if a same output value is observed, the QIF may become different 
depending on which value is assumed to be secret, which is unnatural, and 
(2) a side-channel may exist when further processing is added by system managers 
after getting quantification result. 
Hence, as suggested in \cite{Bi16}, it is better to introduce a new notion
for quantifying dynamic leakage caused by observing a specific output value.
\newline
The contributions of this paper are three-fold.
\begin{itemize}
	\item We present our criteria for an appropriate definition of dynamic leakage and 
	propose two notions that satisfy those criteria.
	We propose two notions because there is a trade-off between
	the easiness of calculation and the preciseness (see Section \ref{sec:2}).
	\item Complexity of computing the proposed dynamic leakages is analyzed 
	for three classes of Boolean programs.
	\item By applying model counting of logical formulae, 
	a prototype was implemented and feasibility of computing those leakages is discussed
	based on experimental results. 
\end{itemize}
According to \cite{Bi16}, we arrange three criteria 
that a `good' definition of dynamic leakage should satisfy, namely, the measure should be 
(R1) non-negative, 
(R2) independent of a secret value to prevent a side channel and
(R3) compatible with existing notions to keep the consistency within QIF 
as a whole (both dynamic leakage and normal QIF). 
Based on those criteria, we come up with two notions of dynamic leakage QIF1 and QIF2, 
where both of them satisfy all (R1), (R2) and (R3). 
QIF1, motivated by entropy-based approach, 
takes the difference between the initial and remaining self-information of the secret 
before and after observing output as dynamic leakage. 
On the other hand, QIF2 models that of the joint probability between secret and output. 
Because both of them are useful in different scenarios, 
we studied these two models in parallel in the theoretical part of the paper. 
We call the problems of computing QIF1 and QIF2 for Boolean programs 
CompQIF1 and CompQIF2, respectively.
For example, we show 
that even for deterministic loop-free programs with uniformly distributed input, 
both CompQIF1 and CompQIF2 are $\sharp P$-hard. 
Next, we assume that secret inputs of a program are uniformly distributed and consider the following method of computing QIF1 and QIF2 (only for deterministic programs for QIF2 by the technical reason mentioned in Section 4): (1) translate a program into a Boolean formula that represents relationship among values of variables during a program execution, (2) augment additional constraints that assign observed output values to the corresponding variables in the formula, 
(3) count models of the augmented Boolean formula projected on secret variables, and (4) calculate the necessary probability and dynamic leakage using the counting result. Based on this method, we conducted experiments using our prototype tool with benchmarks taken from QIF related literatures, in which programs are deterministic, to examine the feasibility of automatic calculation. We also give discussion, in subsection 5.3, on difficulties and possibilities to deal with more general cases, such as, of probabilistic programs. In step (3), we can flexibly use any off-the-shelf model counter. To investigate the scalability of this method, we used four state-of-the-art counters, SharpCDCL \cite{KMM13} and GPMC \cite{SHS17,GPMC} for SAT-based counting, an improved version of aZ3 \cite{PM15} for SMT-based counting, and DSharp-p \cite{MMBH12,DSharp-p} for SAT-based counting in d-DNNF fashion. Finally, we discuss the feasibility of automatic calculation of the leakage in general case.
\medskip\par\noindent
{\bf Related work}
The very early work on \textit{computational complexity} of QIF 
is that of Yasuoka and Terauchi. 
They proved that even the problem of comparing QIF of two programs, 
which is obviously not more difficult than calculating QIF, 
is not a $k$-safety property for any $k$ \cite{YT10}. 
Consequently, self-composition, a successful technique to verify non-interference property, 
is not applicable to the comparison problem. 
Their subsequent work \cite{YT11} proves a similar result for bounding QIF, 
as well as the $PP$-hardness of precisely quantifying QIF 
in all entropy-based definitions for loop-free Boolean programs. 
Chadha and Ummels \cite{CU12} show that 
the QIF bounding problem of recursive programs is not harder than checking reachability
for those programs. 
Despite given those evidences about the hardness of calculating QIF, 
for this decade, \textit{precise QIF analysis} gathers much attention of the researchers. 
In \cite{KMM13}, Klebanov et al. reduce QIF calculation to 
$\sharp$SAT problem projected on a specific set of variables 
as a very first attempt to tackle with automating QIF calculation. 
On the other hand, Phan et al. reduce QIF calculation to $\sharp$SMT problem 
for utilizing existing SMT (satisfiability modulo theory) solver. 
Recently, Val et al. \cite{VEBAH16} reported a method
that can scale to programs of 10,000 lines of code but still based on SAT solver and symbolic execution. 
However, there is still a gap between such improvements and practical use, and 
researchers also work on \textit{approximating QIF}. 
K\"{o}pf and Rybalchenko \cite{KR10} propose approximated QIF computation
by sandwiching the precise QIF by lower and upper bounds 
using randomization and abstraction, respectively with a provable confidence. 
LeakWatch of Chothia et al. \cite{CKN14}, also give approximation with provable confidence 
by executing a program multiple times. 
Its descendant called HyLeak \cite{BKLT17} combines the randomization strategy of its ancestor 
with precise analysis. 
Also using randomization but in Markov Chain Monte Carlo (MCMC) manner, 
Biondi et al. \cite{BEHLMQ18} utilize ApproxMC2, 
an existing model counter created by some of the co-authors. 
ApproxMC2 provides approximation on the number of models of a Boolean formula in CNF 
with adjustable precision and confidence. 
ApproxMC2 uses hashing technique to divide the solution space into smaller buckets 
with almost equal number of elements, 
then count the models for only one bucket and multiply it by the number of buckets. 
As for \textit{dynamic leakage}, McCamant et al. \cite{ME08} 
consider QIF as network flow through programs
and propose a dynamic analysis method that can work with executable files. 
Though this model can scale to very large programs, its precision is relatively not high.
Alvim et al. \cite{ACM16} give some axioms for a reasonable definition of QIF to satisfy
and discuss whether some definitions of QIF satisfy the axioms. Note that these axioms
are for {\em static} QIF measures, which differ from dynamic leakage. However, given a
similarity between static and dynamic notions, we investigated how our new dynamic notions
fit in the lens of the axioms (refer to Section \ref{sec:2}).

\textit{Dynamic information flow analysis} (or taint analysis) is a bit confusing
term that does not mean an analysis of dynamic leakage, but 
a runtime analysis of information flow. 
Dynamic analysis can abort a program as soon as an unsafe information flow is detected. 
Also, hybrid analysis has been proposed for improving dynamic analysis
that may abort a program too early or unnecessarily. 
In hybrid analysis, the unexecuted branches of a program is statically analyzed 
in parallel with the executed branch. 
Among them, Bielova et al. \cite{BBJ16} define the knowledge $\kappa(z)$ of a program variable $z$
as the information on secret that can be inferred from $z$
(technically, $\kappa(z)^{-1}(v)$ is the same of the pre-image of an observed value $v$ of $z$, defined in Section 2). 
In words, hybrid analysis updates the `dynamic leakage' under the assumption that 
the program may terminate at each moment. 
Our method is close to \cite{BBJ16} in the sense that the knowledge $\kappa(z)^{-1}(v)$ is computed. 
The difference is that we conduct the analysis after the a program is terminated
and $v$ is given. 
We think this is not a disadvantage compared with hybrid analysis
because the amount of dynamic leakage of a program is not determined 
until a program terminates in general. 
\medskip\par\noindent
{\bf Structure of the remaining parts:}
Section 2 is dedicated to introduce new notions, 
i.e., QIF1 and QIF2, of dynamic leakage and some properties of them. 
The computational complexity of CompQIF1 and CompQIF2 is discussed in Section 3. 
Section 4 gives details on calculating dynamic leakage based on model counting. 
Experimental results and discussion are provided in Section 5 and the paper is concluded in Section 6.

\section{New Notions for Dynamic Leakage}\label{sec:2}

The standard notion for static quantitative information flow (QIF) is defined as 
the mutual information between random
variables $S$ for secret input and $O$ for observable output:
\begin{equation} 
\QIF = H(S) - H(S|O)
\label{eq:QIF}
\end{equation}
where 
$H(S)$ is the entropy of $S$ and 
$H(S|O)$ is the expected value of $H(S|o)$, 
which is the conditional entropy of $S$ when observing an output $o$. 
Shannon entropy and min-entropy are often used as the definition of entropy, and 
in either case, $H(S)-H(S|O)\ge 0$ always holds by the definition.

In \cite{Bi16}, the author discusses the appropriateness of the 
existing measures for dynamic QIF and points out their drawbacks, 
especially, each of these measures may become negative. 
Hereafter, let ${\cal S}$ and ${\cal O}$ denote the finite sets of input values and output values, 
respectively. 
Since $H(S|O) = \sum_{o\in {\cal O}} p(o) H(S|o)$, 
\cite{Bi16} assumes the following measure
obtained by replacing $H(S|O)$ with $H(S|o)$ in (\ref{eq:QIF}) 
for dynamic QIF:
\begin{equation}
\QIFdyn(o) = H(S)-H(S|o).
\label{eq:QIFdyn}
\end{equation}
However, $\QIFdyn(o)$ may become negative even if a program is deterministic
(see Example \ref{ex:1}). 
Another definition of dynamic QIF is proposed in \cite{CMS09} as
\begin{equation}
\BEL(\dot{s},o) = D_{KL}(p_{\dot{s}}||p_S) - D_{KL}(p_{\dot{s}}||p_{S|o})
\label{eq:belief}
\end{equation}
where $D_{KL}$ is KL-divergence defined as 
$D_{KL}(p||q)= \sum_{s\in {\cal S}}p(s) \log \frac{p(s)}{q(s)}$, and 
$p_{\dot{s}}(s)=1$ if $s=\dot{s}$ and $p_{\dot{s}}(s)=0$ otherwise. 
Intuitively, 
$\BEL(\dot{s},o)$ represents how closer the belief of an attacker approaches 
to the secret $\dot{s}$ by observing $o$. 
For deterministic programs, $\BEL(\dot{s},o) = - \log p(o) \ge 0$ \cite{Bi16}. 
However, $\BEL$ may still become negative if a program is probabilistic
(see Example \ref{ex:2}).
\medskip\par
Let $P$ be a program with secret input variable $S$ 
and observable output variable $O$. 
For notational convenience, we identify 
the names of program variables with the corresponding 
random variables. Throughout the paper, we assume that a program always terminates. 
The syntax and semantics of programs assumed in this paper will be given 
in the next section. 
For $s\in {\cal S}$ and $o\in {\cal O}$, let 
$p_{SO}(s,o)$, $p_{O|S}(o|s)$, $p_{S|O}(s|o)$, $p_S(s)$, $p_O(o)$ denote 
the joint probability of $s\in {\cal S}$ and $o\in {\cal O}$, 
the conditional probability of $o\in {\cal O}$ given $s\in {\cal S}$ (the likelihood), 
the conditional probability of $s\in {\cal S}$ given $o\in {\cal O}$ (the posterior probability), 
the marginal probability of $s\in {\cal S}$ (the prior probability) and 
the marginal probability of $o\in {\cal O}$, respectively.
We often omit the subscripts as $p(s,o)$ and $p(o|s)$ if they are clear from the context.
By definition, 
\begin{eqnarray}
\label{eq:Bayes}
p(s,o) & = & p(s|o)p(o)=p(o|s)p(s), \\
p(o) & = & \sum_{s\in {\cal S}} p(s,o), \\
p(s) & = & \sum_{o\in {\cal O}} p(s,o). 
\end{eqnarray}

We assume that (the source code of) $P$ and 
the prior probability $p(s)$ ($s\in {\cal S}$) are known to an attacker.
For $o\in {\cal O}$, let $\pre_P(o) = \{ s\in {\cal S} \mid p(s|o) > 0 \}$, which is 
called the pre-image of $o$ (by the program $P$). 

\begin{figure}[h]
	\begin{center}\epsfig{file=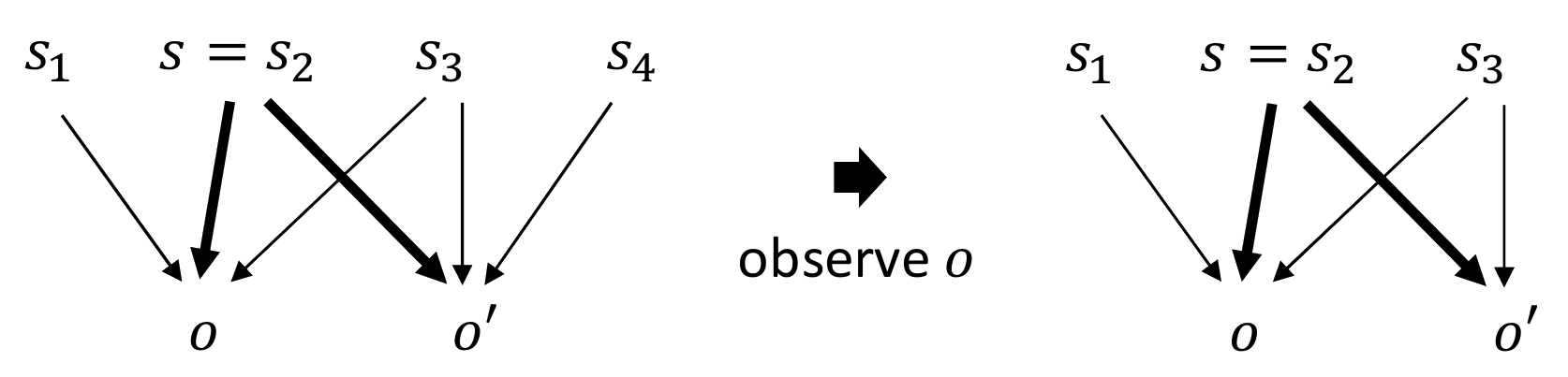,scale=0.4}\end{center}
	\begin{center}\epsfig{file=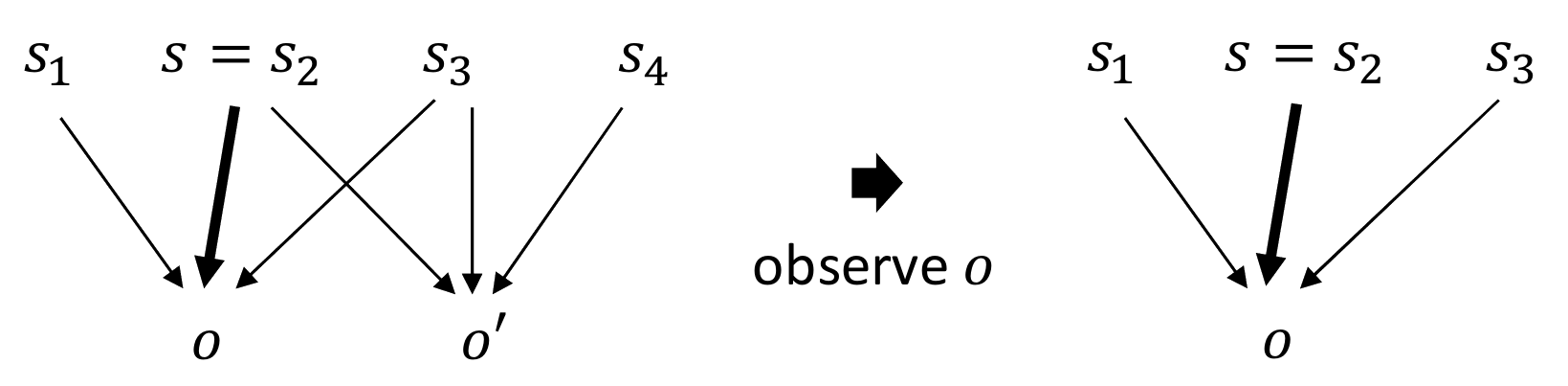,scale=0.4}\end{center}
	\caption{$\QIFone$ (the upper) and $\QIFtwo$ (the lower)}
	\label{fig:1}
\end{figure}

Considering the discussions in the literature, 
we aim to define new notions for dynamic QIF that satisfy the following requirements:
\begin{enumerate}
	\def\labelenumi{(R\arabic{enumi})}
	\def\theenumi{(R\arabic{enumi})}
	\item Dynamic QIF should be always non-negative because
	an attacker obtains some information (although sometimes very small or even zero) when 
	he observes an output of the program.
	\item It is desirable that dynamic QIF is independent of a secret input $s\in {\cal S}$. 
	Otherwise, the controller of the system may change the behavior for protection
	based on the estimated amount of the leakage that depends on $s$, 
	which may be a side channel for an attacker. 
	\item The new notion should be compatible with the existing notions when 
	we restrict ourselves to special cases such as deterministic programs, 
	uniformly distributed inputs, and taking the expected value. 
\end{enumerate}
%
The first proposed notion is the self-information
of the secret inputs consistent with an observed output $o\in {\cal O}$.
Equivalently,
the attacker can narrow down the possible secret inputs
after observing $o$ to the pre-image of $o$ by the program. 
We consider the self-information of $s \in \mathcal{S}$ after 
the observation as the probability of $s$ divided by the sum 
of the probabilities of the inputs in the pre-image of $o$
(see the upper part of Fig. \ref{fig:1}).
\begin{eqnarray}
\QIFone(o) 
& = & - \log (\sum_{s' \in \pre_P(o)} p(s')). \label{eq:QIF1}
\end{eqnarray}
The second notion is
the self-information of
the joint events $s'\in {\cal S}$ and an observed output $o\in {\cal O}$
(see the lower part of Fig. \ref{fig:1}).
This is equal to the the self-information of $o$.
\begin{eqnarray}
\QIFtwo(o) & = & - \log (\sum_{s'\in {\cal S}}p(s',o)) \\
= - \log p(o) & = & - \log p(s,o) + \log p(s|o). \label{eq:QIF2}
\end{eqnarray}
%
Both notions are defined by considering how much
possible secret input values are reduced by observing an output.
We propose two notions because there is a trade-off between
the easiness of calculation and the appropriateness.
As illustrated in Example \ref{ex:2}, $\QIFtwo$ can represent
the dynamic leakage more appropriately than $\QIFone$ in some cases.
On the other hand, the calculation of
$\QIFone$ is easier than $\QIFtwo$
as discussed in Section \ref{sec:4}.
Both notions are independent of the secret input $s\in {\cal S}$
(Requirement (R2)).
\begin{eqnarray}
0 \le \QIFone(o) \le \QIFtwo(o). 
\label{eq:QIF1andQIF2}
\end{eqnarray}
If we assume Shannon entropy, 
\begin{eqnarray}
\label{eq:QIF-a}
\QIF
& = & - \sum_{s\in {\cal S}}{p(s)\log p(s)}  \nonumber \\
&   & \quad + \sum_{o\in {\cal O}} {p(o) \sum_{s\in {\cal S}} {p(s|o)\log p(s|o) }}\\
& = & - \sum_{s\in {\cal S}}{p(s)\log p(s)} \nonumber \\
&   & \quad + \sum_{s\in {\cal S}, o\in {\cal O}} {p(s,o)\log p(s|o) }. 
\label{eq:QIF-b}
\end{eqnarray}
If a program is deterministic, 
for each $s\in {\cal S}$, there is exactly one $o_s\in {\cal O}$ 
such that 
$p(s,o_s)=p(s)$ and $p(s,o)=0$ for $o\not=o_s$, and therefore
\begin{eqnarray}
\label{eq:QIF-d}
\QIF & = & \sum_{s\in {\cal S},o\in {\cal O}}
{p(s,o)(- \log p(s,o) + \log p(s|o))}.
\end{eqnarray}
Comparing (\ref{eq:QIF2}) and (\ref{eq:QIF-d}), we see that 
$\QIF$ is the expected value of $\QIFtwo$, which suggests 
the compatibility of $\QIFtwo$ with $\QIF$ (Requirement (R3))
when a program is deterministic. 
Also, if a program is deterministic, 
$\BEL(\dot{s},o) = - \log p(o)$, which coincides with $\QIFtwo(o)$ (Requirement (R3)). 
By (\ref{eq:QIF1andQIF2}), Requirement (R1) is satisfied. Also in (\ref{eq:QIF1andQIF2}),
$\QIFone(o)=\QIFtwo(o)$ holds for every $o\in {\cal O}$ if and only if 
the program is deterministic. 

\begin{theorem}
	\label{th:det-quant}
	If a program $P$ is deterministic, 
	for every $o\in{\cal O}$ and $s\in{\cal S}$, 
	\[
	\BEL(s,o) = \QIFone(o) = \QIFtwo(o) = - \log p(o).
	\]
	If input values are uniformly distributed, $\QIFone(o) = $
	$\log \frac{|{\cal S}|}{|\pre_P(o)|}$ for every $o\in {\cal O}$. 
	\qed
\end{theorem}

Let us get back to the Example \ref{ex1.1} in the previous section to see how new notions convey
the intuitive meaning of dynamic leakage. We assume: both $source$ and
$output$ are 8-bit numbers of which values are in $0..255$, $source$ is uniformly distributed over this range.
Then, because the program in this example is deterministic, as mentioned above $\QIFone$ coincides with $\QIFtwo$.
We have $\QIFone(output=8)=-\log\frac{241}{256}=0.087 bits$ while $\QIFone(output=o)=-\log\frac{1}{256}=8 bits$ for every $o$ between 9 and 23.
This result addresses well the problem of failing to differentiate vulnerable output from safe ones of QIF.

\begin{example}
	\label{ex:1}
	Consider the following program taken from Example 1 of \cite{Bi16}:
	\begin{quote}
		if $S=s_1$ then $O \gets a$ else $O \gets b$
	\end{quote}
	%
	%
	Assume that the probabilities of inputs are 
	$p(s_1) = 0.875$, $p(s_2) = 0.0625$ and $p(s_3) = 0.0625$. 
	Then, we have the following output and posterior probabilities:
	\begin{quote}
		$p(a)=0.875, p(b)=0.125$\\
		$p(s_1|a)=1, p(s_2|a)=p(s_3|a)=0$\\
		$p(s_1|b)=0, p(s_2|b)=p(s_3|b)=0.5$
	\end{quote}
	If we use Shannon entropy, 
	$H(S) = 0.67$, $H(S|a)=0$ and $H(S|b)=1$. 
	Thus, $\QIFdyn(b) = - 0.33$, which is negative
	as pointed out in \cite{Bi16}.
	Also, 
	$\QIFtwo(a) = - \log p(a) = - \log 0.875=0.19$ and 
	$\QIFtwo(b) = -\log p(b) = - \log 0.125=3$.  
	$\QIFtwo(a) < \QIFtwo(b)$ reflects the fact that
	the difference of the posterior and the prior of each input when observing $b$
	is larger ($s_1: 0.875 \to 0$, $s_2,s_3: 0.0625 \to 0.5$)
	than observing $a$ ($s_1: 0.875 \to 1$, $s_2,s_3: 0.0625 \to 0$).
	
	Since the program is deterministic, 
	$\BEL(s,o) = \QIFone(o) = \QIFtwo(o)$. 
	\[
	\begin{array}{c|cc}
	o          & a         & b \\ \hline
	\QIFdyn(o) & 0.67      &  -0.33 \\
	\QIFtwo(o) & 0.19      &   3
	\end{array}
	\]
	\qed
\end{example}
\begin{example}
	\label{ex:2}
	The next program is quoted from Example 2 of \cite{Bi16} where 
	$c_{1~r}[]_{1-r}~c_2$ means that the program chooses $c_1$ with probability $r$
	and $c_2$ with probability $1-r$.
	\begin{quote}
		if $S=s_1$ then $O \gets a~_{0.81}[]_{0.19}O \gets b$
		\newline
		else $O \gets a~_{0.09}[]_{0.91}O \gets b$
	\end{quote}
	%
	Assume that the probabilities of inputs are
	$p(s_1)=0.25$ and $p(s_2)=0.75$. 
	$
	(p(a), p(b)) = (0.25,0.75)
	\begin{pmatrix}
	0.81 & 0.19 \\
	0.09 & 0.91
	\end{pmatrix}
	=(0.27, 0.73)
	$
	and the posterior probabilities are calculated by (\ref{eq:Bayes}) as: 
	\begin{quote}
		$p(s_1|a)=0.75, p(s_2|a)=0.25$\\
		$p(s_1|b)=0.065, p(s_2|b)=0.935$
	\end{quote}
	Let us use Shannon entropy for $\QIFdyn$. 
	As $H(S) = H(S|a) = -0.25\log 0.25 - 0.75\log 0.75$, 
	$\QIFdyn(a) = H(S) - H(S|a) = 0$. 
	As already discussed in \cite{Bi16}, 
	$\QIFdyn(a)=0$ 
	though an attacker may think that 
	$S=s_1$ is more probable by observing $O=a$. 
	For each $o\in \{a,b\}$, 
	$\BEL(s,o)$ takes different values (one of them is negative)
	depending on whether $s=s_1$ or $s_2$ is the secret input. 
	$\QIFtwo(a) = - \log p(a) = - \log 0.27 = 1.89$ and 
	$\QIFtwo(b) = - \log p(b) = - \log 0.73 = 0.45$. 
	$\QIFone(a) = \QIFone(b) = 0$ because the set of possible input values 
	does not shrink whichever $a$ or $b$ is observed. 
	Similarly to Example \ref{ex:1}, 
	$\QIFtwo(a) > \QIFtwo(b)$ reflects the fact that
	the probability of each input when observing $a$
	varies more largely 
	($s_1: 0.25 \to 0.75$, $s_2: 0.75 \to 0.25$)
	than when observing $b$
	($s_1: 0.25 \to 0.065$, $s_2: 0.75 \to 0.935$). 
	In this example, the number $|{\cal S}|$ of input values is just two, but in general, 
	$|{\cal S}|$ is larger and we can expect $|\pre_P(o)|$ is much smaller than $|{\cal S}|$
	and $\QIFone$ serves a better measure for dynamic QIF. 
	
	\[
	\begin{array}{c|cc}
	o 		& a 	& b \\ \hline
	\QIFdyn(o)	& 0	& 0.46 \\
	\BEL(s_1,o)	& 1.58	& -1.94 \\
	\BEL(s_2,o)	&-1.58	& 0.32 \\
	\QIFone(o)	& 0	& 0 \\
	\QIFtwo(o)	& 1.89	& 0.45
	\end{array}
	\]
\end{example}
A program is {\em non-interferent} if for every $o\in {\cal O}$ such that $p(o)>0$ 
and for every $s\in {\cal S}$, $p(o) = p(o|s)$. 
Assume a program $P$ is non-interferent.
By (\ref{eq:Bayes}), $p(s)=p(s|o)$ for every $o\in {\cal O}$ ($p(o)>0$) and $s\in {\cal S}$, then $\QIF$ = 0 by (\ref{eq:QIF-a}). 
If $P$ is deterministic in addition, $p(o)=p(o|s)=1$ for $o\in {\cal O}$ ($p(o)>0$) and $s\in {\cal S}$. 
That is, if a program is deterministic and non-interferent, it has exactly one possible 
output value.
\medskip\par\noindent
{\bf Relationship to the hybrid monitor} Let us see how our notions relate to the knowledge tracking hybrid monitor proposed by Bielova et al. \cite{BBJ16}.
\begin{example}
	\label{ex:4.1}
	Consider the following program taken from Program 5 of \cite{BBJ16}:
	\begin{quote}
		if $h$ then $z$ $\gets$ $x$ + $y$
		\newline
		else $z$ $\gets$ $y$ - $x$;
		\newline
		output $z$
	\end{quote}
	where $h$ is a secret input, $x$ and $y$ are public inputs and $z$ is a public output. 
\end{example}
In \cite{BBJ16}, the knowledge about secret input $h$ carried by public output $z$ is 
$\kappa (z) = \lambda\rho.$\textit{if}$([\![h]\!]_{\rho}, [\![x+y]\!]_{\rho}, [\![y-x]\!]_{\rho})$
where $\rho$ is an initial environment (an assignment of values to $h$, $x$ and $y$) and $[\![e]\!]_{\rho}$ is the evaluation of $e$ in $\rho$. 
If $h=1$, $x=0$ and $y=1$, then $z=1$. 
In \cite{BBJ16}, to verify whether this value of $z$ reveals any information about $h$ in this setting of public inputs (i.e., $x=0$, $y=1$), 
they take 
$\kappa (z)^{-1}(1) = \{\rho|$\textit{if}$([\![h]\!]_{\rho}, [\![x+y]\!]_{\rho}, [\![y-x]\!]_{\rho})=1\} = \{\rho|$\textit{if}$([\![h]\!]_{\rho}, 1, 1)=1\}$. 
Because \textit{if}$([\![h]\!]_{\rho}, 1, 1)=1$ for every $\rho$, 
\cite{BBJ16} concluded that $z=1$ in that setting leaks no information.
\newline
On the other hand, with that settings of $x=0$ and $y=1$, given $z=1$ as the observed output, $h$ can be either $true$ or $false$. For the program is deterministic, $\QIFone(z=1) = \QIFtwo(z=1) = -\log \sum_{p(s'|o)>0} p(s') = -\log (p(h=true) + p(h=false)) = -\log 1 = 0$, 
which is consistent with that of \cite{BBJ16} though the approach looks different. 
Actually, the function $\kappa (z)$ encodes all information revealed 
from a value of $z$ about secret input. 
By applying $\kappa (z)^{-1}$ for a specific value $o$ of $z$, 
we get the pre-image of $o$. 
In other words, $\kappa (z)^{-1}(o)$ is exactly 
what we are getting toward quantifying our notions of dynamic leakage. 
The monitor proposed in \cite{BBJ16} tracks the knowledge about secret input 
carried by all variables along an execution of a program according to the 
inlined operational semantics. 
It seems, however, impractical to store all the knowledge during an execution, 
and furthermore, 
it would take time to compute the inverse of the knowledge when 
an observed output is fixed.
\medskip\par
The three requirements (R1), (R2) and (R3) we presented summarize the intuitions about 
dynamic leakage following the spirit of \cite{Bi16}. However, those requirements
lack of a firm back-up theory, whilst in \cite{ACM16} Alvim et al. provide a set of axioms 
for QIF. Despite there is difference between QIF and dynamic leakage, we investigated how 
well our notions fit in the lens of those axioms to confirm their feasibility to be used 
as a metric. For the limitations of space, we will skip detailed explanation for the quite 
trivial results.
\newline
(1) For prior vulnerability, both $\QIFone$ and $\QIFtwo$ satisfy \textit{CONTINUITY}, 
\textit{CONVEXITY} (also the loosen version \textit{Q-CONVEXITY}).
\newline
(2) For posterior vulnerability, exactly speaking, we cannot construct the channel matrix $C$, 
because dynamic leakage is about one specific output value but the matrix is for all possibilities. 
Hence, conceptually, those axioms are not applicable in this context. However, by capturing the 
intuitive interpretation of the axioms, we made a small modification (i.e., to use 
$\mathbb{D}\mathcal{Y}\rightarrow\mathbb{R}^{+}$ in stead of $\mathbb{D}^{2}\mathcal{X}\rightarrow\mathbb{R}^{+}$ 
as the type of posterior vulnerability) to investigate the new notions under the spirit of axioms.
So, by definitions above and the meaning of posterior vulnerability in terms of \cite{ACM16}, 
we have $\frac{p(s)}{\sum_{s' \in \pre_P(o)} p(s')}$ and $p(s|o)$ are respectively formulae for
$\widehat{\mathbb{V}}(o)$ in the context of $\QIFone$ and $\QIFtwo$. Given this modification,
we found that $\QIFone$ satisfies all the three axioms: \textit{NI} (Non-Interference), \textit{MONO} (Monotonicity) 
and \textit{DPI} (Data Processing Inequality) whilst $\QIFtwo$ satisfies only the first two axioms but the last one,
\textit{DPI}. In fact, $\QIFtwo$ still aligns well to \textit{DPI} in cases for deterministic programs, and only misses for
probabilistic ones. Please recall that in deterministic cases, $\QIFone \equiv \QIFtwo$
by Theorem \ref{th:det-quant}. Hence, for deterministic programs, $\QIFtwo$ satisfies \textit{DPI} because $\QIFone$ does. 
For it is quite trivial and the space is limited, we will omit the proof of those satisfaction. 
In stead, we will give a counterexample to show that $\QIFtwo$ does not satisfy \textit{DPI} when programs are probabilistic. 
Let $P_1:\{s_1, s_2\}\rightarrow \{u_1, u_2\}$ and $P_2:\{u_1, u_2\}\rightarrow \{v_1\}$ in which $P_2$ is a post-process of $P_1$. 
Also assume the following probabilities: $p(s_1)=p(s_2)=0.5, p(u_1|s_1)=0.1, p(u_2|s_1)=0.9, p(u_1|s_2)=0.3, p(u_2|s_2)=0.7$
and $p(v_1|u_1)=p(v_1|u_2)=1$, in which $s_1,s_2,u_1,u_2,v_1$ annotate events that the corresponding variables have those values.
Given these settings, we have $\widehat{\mathbb{V}}(u_1)=p(s_1|u_1)=\frac{0.5\times 0.1}{0.5\times 0.1 + 0.5\times 0.3}=0.25$,
and $\widehat{\mathbb{V}}(v_1)=p(s_1|v_1)=\frac{0.5\times 0.1+0.5\times 0.9}{0.5\times 0.1+0.5\times 0.9+0.5\times 0.3+0.5\times 0.7}=0.5$.
In other words, $\widehat{\mathbb{V}}(v_1)>\widehat{\mathbb{V}}(u_1)$, which is against to \textit{DPI}.
\newline
\indent It turns out that our proposed notions either satisfy straightly or through some adaptive transformation,
i.e., the type of posterior vulnerability, for all the axioms except \textit{DPI}. For \textit{DPI}, we came to 
the conclusion that it is not suitable as a criterion to verify if a dynamic leakage notion is reasonable.
It is because dynamic leakage is about a specific execution path, in which the inequality of \textit{DPI} does
no longer make sense, rather than the average on all possible execution paths. Therefore, it is not counter-intuitive
that $\QIFtwo$ does not satisfy \textit{DPI} for probabilistic programs while $\QIFtwo$ for deterministic programs 
and $\QIFone$ satisfy \textit{DPI}.

\section{Complexity Results}
\subsection{Program model}
\label{subsec:program}
Let $\Bool =\{ \True, \False\}$ be the set of truth values, 
$\Nat = \{1, 2, \ldots\}$ be the set of natural numbers and $\Natz=\Nat\cup\{0\}$. 
Also let $\Rat$ denote the set of rational numbers. 
We assume probabilistic Boolean programs where every variable stores a truth value 
and the syntactical constructs are 
assignment to a variable, conditional, probabilistic choice, 
while loop, procedure call and sequential composition: 
\begin{eqnarray*}
	e & ::= & \True \mid \False \mid X \mid \neg e \mid e \vee e \mid e \wedge e \\
	c & ::= & \mbox{skip}
	\mid X \gets e 
	\mid \mbox{if } e \mbox{ then } c \mbox{ else } c \mbox{ end}\\
	& & 
	\mid c~_{r}[]_{1-r}~c 
	\mid \mbox{while } e \mbox{ do } c \mbox{ end}
	\mid \pi(\vec{e}; \vec{X}) 
	\mid c ; c
\end{eqnarray*}
where 
$X$ stands for a (Boolean) variable, 
$r$ is a constant rational number such that $0\le r\le 1$. 
In the above BNFs, objects derived from the syntactical categories $e$ and $c$
are called expressions and commands, respectively.

A procedure $\pi$ has the following syntax: 
\[ \mbox{in } \vec{X}; \mbox{ out } \vec{Y}; \mbox{ local } \vec{Z}; c \]
where $\vec{X}, \vec{Y}, \vec{Z}$ are sequences of input, output and local variables, 
respectively (which are disjoint from one another). 
Let $Var(\pi) = \{ V \mid V \mbox{ appears in } \vec{X}, \vec{Y} \mbox{ or } \vec{Z} \}$. 
We will use the same notation $Var(e)$ and $Var(c)$ for an expression $e$ and a command $c$. 
A program is a tuple of procedures $P = (\pi_1, \pi_2, \ldots, \pi_k)$ where 
$\pi_1$ is the main procedure. 
$P$ is also written as $P(\vec{S},\vec{O})$ to emphasize the input and output variables 
$\vec{S}$ and $\vec{O}$ of 
$\pi_1 = \mbox{in } \vec{S}; \mbox{ out } \vec{O}; \mbox{ local } \vec{Z}; c.$
%

A command $X \gets e$ assigns the value of Boolean expression $e$ to variable $X$. 
A command $c_{1~r}[]_{1-r}~c_2$ means that the program chooses $c_1$ with probability $r$
and $c_2$ with probability $1-r$.  
Note that this is the only probabilistic command. 
A command $\pi(\vec{e}; \vec{X})$ is a recursive procedure call to $\pi$ 
with actual input parameters $\vec{e}$ and return variables $\vec{X}$. 
The semantics of the other constructs are defined in the usual way. 
%
%

The size of $P$ is 
the sum of the number of commands and 
the maximum number of variables in a procedure of $P$. 

If a program does not have a recursive procedure call and $k=1$, it is called a (non-recursive) while program. 
If a while program does not have a while loop, it is called a loop-free program 
(or straight-line program). 
If a program does not have a probabilistic choice, it is {\em deterministic}. 

\subsection{Assumption and overview}

We define the problems $\CQIFone$ and $\CQIFtwo$ as follows.

\begin{quote}
	Inputs: 
	a probabilistic Boolean program $P$, \\
	\quad an observed output value $o\in {\cal O}$, and \\
	\quad a natural number $j$ (in unary) specifying the error bound. \\
	Problem: Compute $\QIFone(o)$ (resp. $\QIFtwo(o))$ for $P$ and $o$.
\end{quote}
\noindent
(General assumption) 
\begin{enumerate}
	\def\labelenumi{(A\arabic{enumi})}
	\def\theenumi{(A\arabic{enumi})}
	\item
	The answer to the problem $\CQIFone$ (resp. $\CQIFtwo$) should be given
	as a rational number (two integer values representing the numerator and denominator)
	representing the probability $\sum_{s'\in\pre_P(o)} p(s')$ (resp. $p(o)$). 
	\item 
	If a program is deterministic or non-recursive, the answer should be exact.
	Otherwise, the answer should be within $j$ bits of precision, i.e., 
	$| \mbox{ (the answer)} - \sum_{s'\in\pre_P(o)} p(s') (\mbox{resp. } p(o)) ~| \le 2^{-j}$. 
	
\end{enumerate}
\noindent
If we assume (A1), 
we only need to perform additions and multiplications the number of times 
determined by an analysis of a given program, avoiding the computational difficulty
of calculating the exact logarithm. 
The reason for assuming (A2) is that the exact reachability probability 
of a recursive program is not always a rational number even if all the transition
probabilities are rational (Theorem 3.2 of \cite{EY05}). 

When we discuss lower-bounds, 
we consider the corresponding decision problem by adding a candidate answer
of the original problem as a part of an input. 
The results on the complexity of $\CQIFone$ and $\CQIFtwo$ 
are summarized in Table \ref{tbl:1}. 
As mentioned above, if a program is deterministic, $\QIFone = \QIFtwo$. 
\begin{table}[h]
	\caption{Complexity results}
	\begin{center}
		\begin{tabular}{|c|c|c|c|}
			\hline
			\multirow{2}{*}{programs} & \multirow{2}{*}{deterministic} 
			& \multicolumn{2}{c|}{probabilistic} \\ \cline{3-4}
			& & $\CQIFone$ & $\CQIFtwo$ \\ \hline
			\multirow{3}{*}{loop-free} 
			& PSPACE           		& PSPACE          		& PSPACE\\
			& $\sharp P$-hard  		&(Theorem \ref{th:loop-free})	&(Theorem \ref{th:loop-free})\\
			&(Proposition \ref{lm:d-loop-free})	& $\sharp P$-hard 		& $\sharp P$-hard \\ \hline
			\multirow{3}{*}{while}     
			& PSPACE-comp      		& PSPACE-comp     		& EXPTIME \\
			&(Proposition \ref{lm:d-while})&(Theorem \ref{th:while-1})		&(Theorem \ref{th:while-2})\\
			&                  		&                 		& PSPACE-hard\\ \hline
			\multirow{3}{*}{recursive} 
			& EXPTIME-comp     		& EXPSPACE        		& EXPSPACE \\
			&(Proposition \ref{lm:d-recursive})	&(Theorem \ref{th:recursive})	&(Theorem \ref{th:recursive})\\
			&                  		& EXPTIME-hard    		& EXPTIME-hard \\ \hline
		\end{tabular}
	\end{center}
	\label{tbl:1}
\end{table}
%
%

Recursive Markov chain (abbreviated as RMC) is defined in \cite{EY05} 
by assigning a probability to each transition in 
recursive state machine (abbreviated as RSM) \cite{AEY01}. 
Probabilistic recursive program in this paper is similar to RMC except that 
there is no program variable in RMC. 
If we translate a recursive program into an RMC, the number of states 
of the RMC may become exponential to the number of Boolean variables in the recursive program. 
In the same sense, 
deterministic recursive program corresponds to RSM, or equivalently, pushdown systems (PDS) 
as mentioned and used in \cite{CU12}. 
Also, probabilistic while program corresponds to Markov chain. 
We will review the definition of RMC in Section \ref{subsec:RMC}. 

\subsection{Deterministic case}

We first show lower bounds for deterministic loop-free, while and recursive programs. 
For deterministic recursive programs, we give EXPTIME upper bound 
as a corollary of Theorem \ref{th:recursive}. 

\begin{proposition} 
	\label{lm:d-loop-free}
	$\CQIFone (=\CQIFtwo)$ is $\sharp P$-hard for deterministic loop-free programs even if the input values are uniformly distributed.\\
\end{proposition}
	(Proof) 
	We show that $\sharp$SAT can be reduced to $\CQIFone$ where the input values are uniformly distributed. 
	It is necessary and sufficient for $\CQIFone$ to compute the number of inputs $\vec{s}$ such that $p(\vec{s}|\vec{o})> 0$
	because $\sum_{p(\vec{s}|\vec{o})>0}p(\vec{s})=|\{\vec{s}\in \vec{{\cal S}}\mid p(\vec{s}|\vec{o})> 0\}|/|{\vec{{\cal S}}}|$. 
	For a given propositional logic formula $\phi$ with Boolean variables $\vec{S}$, we just construct a program $P$ 
	with input variables $\vec{S}$ and an output variable $O$ such that the value of $\phi$ for $\vec{S}$ is stored to $O$. 
	Then, the result of $\CQIFone$ with $P$ and $o=\top$ coincides with the number of models of $\phi$. 
	\qed

\begin{proposition} 
	\label{lm:d-while}
	$\CQIFone (=\CQIFtwo)$ is PSPACE-hard for deterministic while programs.\\
\end{proposition}
	(Proof)
	The proposition can be shown in the same way as the proof of 
	PSPACE-hardness of 
	the non-interference problem for deterministic while programs 
	by a reduction from quantified Boolean formula (QBF) 
	validity problem given in \cite{CU12} as follows. 
	For a given QBF $\varphi$, 
	we construct a deterministic while program $P$ having one output variable
	such that $P$ is non-interferent if and only if $\varphi$ is valid as
	in the proof of Proposition 19 of \cite{CU12}. 
	The deterministic program is non-interferent if and only if the output of the program is always
	$\top$, i.e., $p(\top)=1$. Thus, we can decide if $\phi$ is valid by checking whether $p(\top)=1$
	or not for the deterministic program, the output value $\top$, and the probability 1.
	\qed

\begin{proposition} 
	\label{lm:d-recursive}
	$\CQIFone (=\CQIFtwo)$ is EXPTIME-complete for deterministic recursive programs.\\
\end{proposition}
	(Proof)
	EXPTIME upper bound can be shown 
	by translating a given program to a pushdown system (PDS). 
	Assume we are given a deterministic recursive program $P$ and an output value $o\in {\cal O}$. 
	We apply to $P$ the translation to a recursive Markov chain (RMC) $A$
	in the proof of Theorem \ref{th:recursive}. 
	The size of $A$ is exponential to the size of $P$. 
	Because $P$ is deterministic, $A$ is also deterministic;
	$A$ is just a recursive state machine (RSM) or equivalently, a PDS.
	It is well-known \cite{BEM97} that the pre-image of a configuration $c$ of a PDS $A$
	$\pre_A(c) = \{ c' \mid c' \mbox{ is reachable to } c \mbox{ in } A \}$
	can be computed in polynomial time by so-called P-automaton construction. 
	Hence, by specifying configurations outputting $o$ as $c$, we can compute 
	$\pre_P(o) = \pre_A(c)$ in exponential time. 
	
	The lower bound can be shown in the same way as the 
	EXPTIME-hardness proof of 
	the non-interference problem for deterministic recursive programs 
	by a reduction from the membership problem for polynomial space-bounded 
	alternating Turing machines (ATM) given in the proof of Theorem 7 of \cite{CU12}.
	From a given polynomial space-bounded ATM $M$ and an input word $w$ to $M$, 
	we construct a deterministic recursive program $P$ having one output variable
	such that $P$ is non-interferent if and only if $M$ accepts $w$ as in \cite{CU12}. 
	As in the proof of Proposition \ref{lm:d-while}, 
	we can reduce to $\CQIFone$ instead of reducing to the non-interference problem. 
	\qed

\subsection{Loop-free programs}

We show upper bounds for loop-free programs.
For $\CQIFtwo$, the basic idea is similar to the one in \cite{CU12}, but 
we have to compute the conditional probability $p(\vec{o}|\vec{s})$. 
For $\CQIFone$, $\sharp P^{NP}$ upper bound can be obtained by a similar
result on model counting if the input values are uniformly distributed. 


\begin{theorem} 
	\label{th:loop-free}
	$\CQIFone$ and $\CQIFtwo$ are solvable in PSPACE 
	for probabilistic loop-free programs.  
	$\CQIFone$ is solvable in $\sharp P^{NP}$ if the input values are uniformly distributed. \\
\end{theorem}
	(Proof)
	We first show that $\CQIFtwo$ is solvable in PSPACE for probabilistic loop-free programs.
	If a program is loop-free, we can compute $p(\vec{o}|\vec{s})$ for every $\vec{s}$ in the same way as in \cite{CU12}, 
	multiply it by $p(\vec{s})$ and sum up in PSPACE.  
	Note that in \cite{CU12}, it is assumed that a program is deterministic and 
	input values are uniformly distributed, 
	and hence it suffices to count the input values $\vec{s}$ such that $p(\vec{o}|\vec{s})=1$, 
	which can be done in $P^{CH3}$. 
	In contrast, we have to compute the sum of the probabilities of $p(\vec{s})p(\vec{o}|\vec{s})$ 
	for all $\vec{s}\in \vec{{\cal S}}$. 
	We can easily see that $\CQIFone$ is solvable in PSPACE for probabilistic loop-free programs 
	in almost the same way as $\CQIFtwo$. Instead of summing up $p(\vec{s})p(\vec{o}|\vec{s})$  for all $\vec{s}\in \vec{{\cal S}}$, 
	we just have to sum up $p(\vec{s})$ for all $\vec{s}\in \vec{{\cal S}}$ such that $p(\vec{o}|\vec{s})>0$ (if and only if $p(\vec{s}|\vec{o})>0$). 
	
	Next, we show that $\CQIFone$ is solvable in $\sharp P^{NP}$ if the input values are uniformly distributed. 
	As stated in the proof of Proposition~\ref{lm:d-loop-free}, in this case, 
	$\CQIFone$ can be solved by computing the number of inputs $s$ such that $p(\vec{s}|\vec{o})> 0$. 
	Deciding $p(\vec{s}|\vec{o})> 0$ for a given probabilistic loop-free program $P$ can be reduced to the satisfiability 
	problem of a propositional logic formula. Note that for any probabilistic choice like $X \gets c_1~_{r}[]_{1-r}~c_2$ with $0<r<1$, 
	we just have to treat it as a non-deterministic choice like $X = c_1~or~X = c_2$ because all we need to know is whether $p(\vec{s}|\vec{o})>0$. 
	We construct from $P$ a formula $\phi$ with Boolean variable corresponding to 
	input and output variables of $P$ and intermediate variables. 
	Here, we abuse the symbols $\vec{S}$ and $\vec{O}$, which are used for 
	the variables of $P$, also as the Boolean variables corresponding 
	to them, respectively. 
	The formula $\phi$ is constructed such that $\phi\land \vec{S}=\vec{s} 
	\land \vec{O} = \vec{o}$ is satisfiable if and only if $p(\vec{s}|\vec{o})$> 0 
	for $\vec{s}$ and $\vec{o}$. 
	Thus, the number of inputs $\vec{s}$ such that $p(\vec{s}|\vec{o})> 0$ is the number of truth assignments for $\vec{S}$ 
	such that $\phi \land \vec{O}=\vec{o}$ is satisfiable, i.e., the number of projected models on $\vec{S}$. 
	This counting can be done in $\sharp P^{NP}$ because projected model counting is in $\sharp P^{NP}$~\cite{ACMS15}. 
	\qed

\subsection{While programs}

We show upper bounds for while programs. 
For $\CQIFone$, we reduce the problem to the reachability problem of a graph representing the 
state reachability relation. 
An upper bound for $\CQIFtwo$ will be obtained as a corollary of Theorem \ref{th:recursive}. 

\begin{theorem} 
	\label{th:while-1}
	$\CQIFone$ is PSPACE-complete for probabilistic while programs.\\
\end{theorem}
	%
	(Proof) It suffices to show that $\QIFone$ is solvable in PSPACE for probabilistic while programs.
	QIF1 for probabilistic while programs is reduced to the reachability problem of graphs 
	that represents the reachability among states of $P$. 
	We construct a directed graph $G$ from a given program $P$ as follows. 
	Each node $(l, \sigma)$ on $G$ uniquely corresponds to a location $l$ on $P$ and 
	an assignment $\sigma$ for all variables in $P$. 
	An edge from $(l, \sigma)$ to $(l', \sigma')$ represents that if the program is running at $l$ with $\sigma$ 
	then, with probability greater than $0$, it can transit to $l'$ with $\sigma'$ by executing the command at $l$. 
	Deciding the reachability from a node to another node can be done in nondeterministic $\log$ space of the size of the graph. 
	The size of the graph is exponential to the size of $P$ due to exponentially many assignments for variables.
	We see that $p(\vec{s}|\vec{o})>0$ if and only if 
	there are two nodes $(l_s, \sigma_s)$ and $(l_o, \rho_o)$ such that 
	$l_s$ is the initial location, $l_o$ is an end location, $\sigma_s(\vec{S})=\vec{s}$, $\sigma_o(\vec{O})=\vec{o}$, 
	and $(l_o, \rho_o)$ is reachable from $(l_s, \sigma_s)$ in $G$. 
	Thus, $p(\vec{s}|\vec{o})>0$ can be decided in PSPACE, 
	and also $\sum_{p(\vec{s}|\vec{o})>0}p(\vec{s})$ can be computed in PSPACE. 
	\qed

\begin{theorem} 
	\label{th:while-2}
	$\CQIFtwo$ is solvable in EXPTIME for probabilistic while programs. \\
\end{theorem}
	(We postpone the proof until we show the result on recursive programs.)
	\qed

\subsection{Recursive programs} 

As noticed in the end of Section \ref{subsec:program}, 
we will use recursive Markov chain (RMC) to give 
upper bounds of the complexity of 
$\CQIFone$ and $\CQIFtwo$ for recursive programs because
RMC has both probability and recursion and 
the complexity of the reachability probability problem for RMC
was already investigated in \cite{EY05}. 

\subsubsection{Recursive Markov chains}
\label{subsec:RMC}

A {\em recursive Markov chain} (RMC) \cite{EY05} is a tuple $A = (A_1, \ldots, A_k)$ where 
each $A_i=(N_i,B_i,Y_i,En_i,Ex_i,\delta_i)$ ($1\le i\le k$) is a {\em component graph} 
(or simply, component) consisting of:
\begin{itemize}
	\item a finite set $N_i$ of {\em nodes}, 
	\item a set $En_i\subseteq N_i$ of {\em entry nodes}, 
	and a set $Ex_i\subseteq N_i$ of {\em exit nodes}, 
	\item a set $B_i$ of {\em boxes}, and a mapping $Y_i: B_i \to \{1,\ldots, k\}$ 
	from boxes to (the indices of) components. 
	To each box $b\in B_i$, 
	a set of {\em call sites} $Call_b = \{ (b,en) \mid en\in En_{Y_i(b)} \}$ and 
	a set of {\em return sites} $Ret_b = \{ (b,ex) \mid ex\in Ex_{Y_i(b)} \}$ are associated. 
	\item $\delta_i$ is a finite set of {\em transitions} of the form $(u,p_{u,v},v)$ where
	\begin{itemize}
		\item the source $u$ is either a non-exit node $u\in N_i\backslash Ex_i$ or a return site.
		\item the destination $v$ is either a non-entry node $v\in N_i\backslash En_i$ or a call site.
		\item $p_{u,v}\in\Rat$ is a rational number between 0 and 1 
		representing the transition probability from $u$ to $v$. 
		We require for each source $u$, 
		$\sum_{\{ v' \mid (u,p_{u,v'},v')\in\delta_i\}}p_{u,v'} = 1$. 
		We write $u \ptrans{p_{u,v}} v$ instead of $(u,p_{u,v},v)$ for readability.
		Also we abbreviate $u \ptrans{1} v$ as $u\to v$. 
	\end{itemize}
\end{itemize}
Intuitively, a box $b$ with $Y_i(b)=j$ denotes an invocation of component $j$ from component $i$. 
There may be more than one entry node and exit node in a component. 
A call site $(b,en)$ specifies the entry node 
from which the execution starts when called from the box $b$.
A return site has a similar role to specify the exit node. 

Let $Q_i = N_i \cup \bigcup_{b\in B_i} {(Call_b \cup Ret_b)}$, 
which is called the set of {\em locations} of $A_i$. 
We also let $N = \bigcup_{1\le i\le k}N_i$, $B = \bigcup_{1\le i\le k}B_i$, 
$Y = \bigcup_{1\le i\le k}Y_i$ where $Y:B \to \{1,\ldots, k\}$, 
$\delta = \bigcup_{1\le i\le k}\delta_i$ and $Q = \bigcup_{1\le i\le k}Q_i$.

The probability $p_{u,v}$ of a transition $u\ptrans{p_{u,v}}v$ is a rational number
represented by a pair of non-negative integers, the numerator and denominator. 
The size of $p_{u,v}$ is the sum of the numbers of bits of these two integers, 
which is called the {\em bit complexity} of $p_{u,v}$.

The semantics of an RMC $A$ is given by the global (infinite state) Markov chain 
$M_A=(V,\Delta)$ induced from $A$ where
$V = B^{\ast} \times Q$ is the set of global states and 
$\Delta$ is the smallest set of transitions satisfying the following conditions:
\begin{enumerate}
	\def\labelenumi{(\arabic{enumi})}
	\def\theenumi{(\arabic{enumi})}
	\item For every $u\in Q$, $(\varepsilon, u)\in V$ where $\varepsilon$ is the empty string.
	\item If $(\alpha,u)\in V$ and $u\ptrans{p_{u,v}}v\in\delta$, 
	then $(\alpha,v)\in V$ and $(\alpha,u)\ptrans{p_{u,v}}(\alpha,v)\in\Delta$.
	\item If $(\alpha,(b,en))\in V$ with $(b,en)\in Call_b$, 
	then $(\alpha b,en)\in V$ and $(\alpha,(b,en))\to (\alpha b,en)\in\Delta$. 
	\item If $(\alpha b, ex)\in V$ with $(b,ex)\in Ret_b$, 
	then $(\alpha,(b,ex))\in V$ and $(\alpha b,ex) \to (\alpha,(b,ex))\in\Delta$. 
\end{enumerate}
Intuitively, $(\alpha,u)$ is the global state where $u$ is a current location and 
$\alpha$ is a pushdown stack, which is a sequence of box names
where the right-end is the stack top. 
(2) defines a transition within a component. 
(3) defines a procedure call from a call site $(b,en)$; the box name $b$ is pushed 
to the current stack $\alpha$ and the location is changed to $en$. 
(4) defines a return from a procedure; the box name $b$ at the stack top is popped 
and the location becomes the return site $(b,ex)$. 
For a location $u\in Q_i$ and an exit node $ex\in Ex_i$ in the same component $A_i$, 
let $q^{\ast}_{(u,ex)}$ denote the probability of reaching $(\varepsilon,ex)$
starting from $(\varepsilon,u)$
\footnote{Though we usually want to know $q^{\ast}_{(en,ex)}$ for an entry node $en$,
	the reachability probability is defined in a slightly more general way.}.  
Also, let $q^{\ast}_u = \sum_{ex\in Ex_i}q^{\ast}_{(u,ex)}$.
The reachability probability problem for RMCs is 
the one to compute $q^{\ast}_{(u,ex)}$ within $j$ bits of precision
for a given RMC $A$, 
a location $u$ and an exit node $ex$ in the same component of $A$ and 
a natural number $j$ in unary.
\newline 
\indent The following property is shown in \cite{EY05}.
\begin{proposition}
	\label{pp:RMC}
	The reachability probability problem for RMCs can be solved in PSPACE.
	Actually, $q^{\ast}_{(u,ex)}$ can be computed for every pair of $u$ and $ex$
	simultaneously in PSPACE by calculating the least fixpoint of the 
	nonlinear polynomial equations induced from a given RMC. 
	\qed
\end{proposition}

\subsubsection{Results}

\begin{theorem}
	\label{th:recursive}
	$\CQIFone$ and $\CQIFtwo$ are solvable in EXPSPACE 
	for probabilistic recursive programs. \\
\end{theorem}
	(Proof) We will prove
	the theorem by translating
	a given program $P$ into a recursive Markov chain (RMC)
	whose size is exponential to the size of $P$. 
	By Proposition \ref{pp:RMC}, we obtain EXPSPACE upper bound. 
	Because an RMC has no program variable, we expand Boolean variables in $P$ to
	all (reachable) truth-value assignments to them. 
	A while command is translated into two transitions; one for exit and the other for while-body.
	A procedure call is translated into a box and transitions connecting to/from the box. 
	For the other commands, the translation is straightforward. 
	
	Let $P=(\pi_1,\ldots,\pi_k)$ be a given program.
	For $1\le i\le k$, 
	let $Val(\pi_i)$ be the set of truth value assignments to $Var(\pi_i)$. 
	We will use the same notation $Val(e)$ and $Val(c)$ for an expression $e$ and a command $c$. 
	For an expression $e$ and an assignment $\theta \in Val(e)$, we write $e\theta$
	to denote the truth value obtained by evaluating $e$ under the assignment $\theta$. 
	For an assignment $\theta$ and a truth value $c$, let $\theta[X\gets c]$ denote
	the assignment identical to $\theta$ except $\theta[X\gets c](X) = c$. 
	We use the same notation for sequences of variables $\vec{X}$ and truth values $\vec{c}$
	as $\theta[\vec{X}\gets \vec{c}]$. 
	
	We construct the RMC $A=(A_1,\ldots,A_k)$ from $P$ where each component graph
	$A_i=(N_i,B_i,Y_i,En_i,Ex_i,\delta_i)$ ($1\le i\le k$) is constructed from 
	$\pi_i = \mbox{in } \vec{X}; \mbox{ out } \vec{Y}; \mbox{ local } \vec{Z}; c_i$ as follows.
	\begin{itemize}
		\item
		$En_i = \{ (c_i,\theta) \mid \theta\in Val(\pi_i)
		\mbox{ where } \theta(W) \mbox{ is arbitrary for}\\ W\in\vec{X} \mbox{ and }
		\theta(W)=\bot \mbox{ for } W\in\vec{Y}\cup\vec{Z} \}$.
		\item
		$Ex_i = \{ \sigma \mid \sigma \mbox{ is an assignment to } \vec{Y} \}$.
		\item
		$N_i$, $B_i$, $Y_i$ and $\delta_i$ are constructed as follows.
	\end{itemize}
	\begin{enumerate}
		\def\labelenumi{(\arabic{enumi})}
		\def\theenumi{(\arabic{enumi})}
		\item $N_i \gets En_i$, 
		$B_i \gets \emptyset$, 
		$Y_i \gets$ the function undefined everywhere, 
		$\delta_i \gets \{ (\mbox{skip},\theta) \to \theta|_{\vec{Y}} \mid \theta \in Val(\pi_i) \}$
		where $\theta|_{\vec{Y}}$ is the restriction of $\theta$ to $\vec{Y}$.
		Note that $\theta|_{\vec{Y}}\in Ex_i$. 
		\item Repeat the following construction until all the elements in $N_i$ are marked:\\
		Choose an unmarked $(c,\theta)$ from $N_i$, mark it and do one of the followings
		according to the syntax of $c$. 
		\begin{enumerate}
			\def\labelenumii{(\roman{enumii})}
			\def\theenumii{(\roman{enumii})}
			\item 
			$c = X \gets e ; c'$. 
			Add $(c', \theta[X\gets e\theta])$ to $N_i$ and
			add $(c,\theta)\to(c', \theta[X\gets e\theta])$ to $\delta_i$.
			\item 
			$c = \mbox{if } e \mbox{ then } c_1 \mbox{ else } c_2 \mbox{ end} ; c'$. 
			Add $(c_1;c',\theta)$ to $N_i$ and 
			add $(c,\theta) \to (c_1;c',\theta)$ to $\delta_i$ 
			if $e\theta=\top$. 
			Add $(c_2;c',\theta)$ to $N_i$ and 
			add $(c,\theta) \to (c_2;c',\theta)$ to $\delta_i$ 
			if $e\theta=\bot$. 
			\item
			$c = c_{1~r}[]_{1-r}~c_2 ; c'$. 
			Add $(c_1;c',\theta)$ and $(c_2;c',\theta)$ to $N_i$.
			Add $(c,\theta)\ptrans{r}(c_1;c',\theta)$ and $(c,\theta)\ptrans{1-r}(c_2;c',\theta)$
			to $\delta_i$. 
			\item
			$c = \mbox{while } e \mbox{ do } c_1 \mbox{ end} ; c'$. 
			Add $(c',\theta)$ to $N_i$ and 
			add $(c,\theta) \to (c',\theta)$ to $\delta_i$ if $e\theta = \bot$. 
			Add $(c_1;c, \theta)$ to $N_i$ and 
			add $(c,\theta) \to (c_1;c, \theta)$ to $\delta_i$ 
			if $e\theta = \top$. 
			\item 
			$c = \pi_j(\vec{e'}; \vec{X'}) ; c'$ 
			where
			$\pi_j = $\\ $\mbox{in } \vec{X''}; \mbox{ out } \vec{Y''}; \mbox{ local } \vec{Z''}; c_j$. 
			Define $Y_i(b) = j$. 
			Add a new box $b$ to $B_i$. 
			Add $(c,\theta) \to (b,(c_j, \bot[\vec{X''}\gets\vec{e'}\theta]))$ to $\delta_i$ where
			the assignment $\bot$ denotes the one that assigns $\bot$ to every variable. 
			For every $\sigma\in Ex_j$, 
			\begin{quote}
				add $(c', \theta[\vec{X'}\gets\vec{Y''}\sigma])$ to $N_i$ and 
				add $(b,\sigma) \to (c', \theta[\vec{X'}\gets\vec{Y''}\sigma])$ to $\delta_i$
				(see Fig. \ref{fig:2}).%
				\begin{figure}[h]
					\begin{center}\epsfig{file=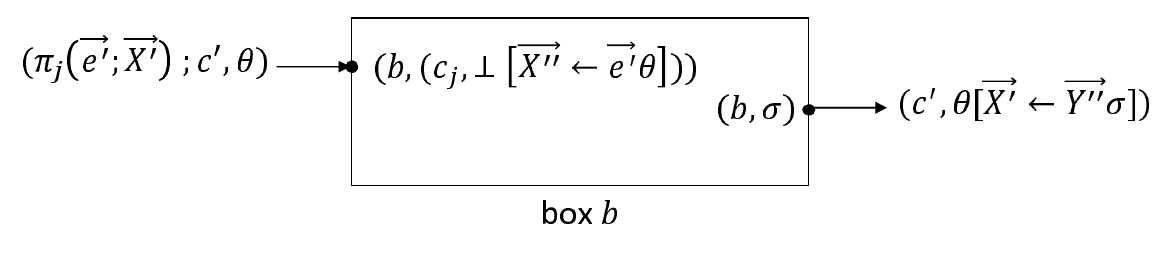,scale=0.7}\end{center}
					\caption{Construction of an RMC from a recursive program}
					\label{fig:2}
				\end{figure}
			\end{quote}
		\end{enumerate}
	\end{enumerate}
	The number $|Q|$ of locations of the constructed RMC $A$ is exponential to the size of $P$. 
	More precisely, $|Q|$ is in the order of 
	the number of commands in $P$ multiplied by $2^N$ where $N$ is the maximum number of
	variables appearing in a procedure of $P$ because we construct locations of $A$
	by expanding each variable to two truth values. 
	Recall that 
	both $\QIFone(o)$ and $\QIFtwo(o)$ can be computed by calculating $p(o|s')$ for each $s'\in {\cal S}$, 
	i.e., the reachability probability from $s'$ to $o$.  
	By Proposition \ref{pp:RMC}, the reachability probability problem for RMCs are in PSPACE, and 
	hence $\CQIFone$ and $\CQIFtwo$ are solvable in EXPSPACE. 
	\qed

\subsubsection*{Proof of Theorem \ref{th:while-2}}


Let $P$ be a given probabilistic while program and $o\in {\cal O}$ is an output value.
Our algorithm works as follows.
\begin{enumerate}
	\item Compute $\pre_P(o)$. 
	\item Calculate $\sum_{s'\in {\cal S}}p(s')p(o|s')$ \quad (see (\ref{eq:QIF2})).
\end{enumerate}
In the proof of Theorem \ref{th:recursive}, a given program $P$ 
is translated into a recursive Markov chain $A$ whose size is exponential to the size of $P$. 
If a given program $P$ is a while program, 
$A$ is an ordinary (non-recursive) Markov chain. 
The constraint on the stationary distribution vector of $A$ is 
represented by 
a system of linear equations whose size is polynomial of the size of $A$ (see \cite{MU05} for example)
and the system of equations can be solved in polynomial time. 
Hence $\CQIFtwo$ is solvable in EXPTIME. 

\section{Model counting-based computation of dynamic leakage} \label{sec:4}
In the previous section, we show that the problems of calculating dynamic leakage, 
i.e., $\CQIFone$ and $\CQIFtwo$, are computationally hard. 
We still, however, propose a practical solution to these problems 
by reducing them to model counting problems. 
\medskip\par\noindent
{\bf Reduction to model counting}
Model counting 
is a well-known and powerful technique in quantitative software analysis and verification including QIF analysis. 
In existing studies, QIF calculation has been reduced to model counting of a logical formula 
using SAT solver \cite{KMM13} or SMT solver \cite{PM15}. 
Similarly, we are showing that it is possible to reduce 
$\CQIFone$ and $\CQIFtwo$ to model counting in some reasonable assumptions. 
Let us consider what is needed to compute 
based on their definitions (\ref{eq:QIF1}) and (\ref{eq:QIF2}), 
i.e., $\QIFone = -\log (\sum_{p(s'|o)>0} p(s'))$ 
and $\QIFtwo = -\log (\sum_{s'\in S}	p(s')p(o|s'))$.

For calculating $\QIFone$ for a given output value $o$, it suffices 
(1) to enumerate input values $s'$ that satisfy $p(s'|o) > 0$ (i.e., possible to produce $o$), and 
(2) to sum the prior probabilities over the enumerated input values $s'$. 
(2) can be computed from the prior probability distribution of input values, 
which is reasonable to assume. When input values are uniformly distributed, 
only step (1) is needed because $\QIFone$ is simplified to
$\log\frac{|{\cal S}|}{|\pre_P(o)|}$ by Theorem \ref{th:det-quant}.

Let us consider $\QIFtwo$. 
For {\it deterministic} programs, $\QIFone = \QIFtwo$ holds
(Theorem \ref{th:det-quant}). 
For {\it probabilistic} programs, 
we need to compute the conditional probability $p(o|s')$ for each $s'$, 
meaning that we have to examine all possible execution paths. 
We would leave $\CQIFtwo$ for probabilistic programs as future work. 

Given a program $P$ together with its prior probability distribution on input, and an observed output $o$, all we need for $\CQIFone$ and $\CQIFtwo$ (deterministic case for the latter)
is the enumeration of $\pre_P(o)$, the input values consistent with $o$. Also, we can forget the probability of a choice command and
regard it just as a nondeterministic choice. Especially when input values are uniformly distributed, only the number of elements of $\pre_P(o)$ is needed.

\begin{figure}[h]
	\begin{center}\epsfig{file=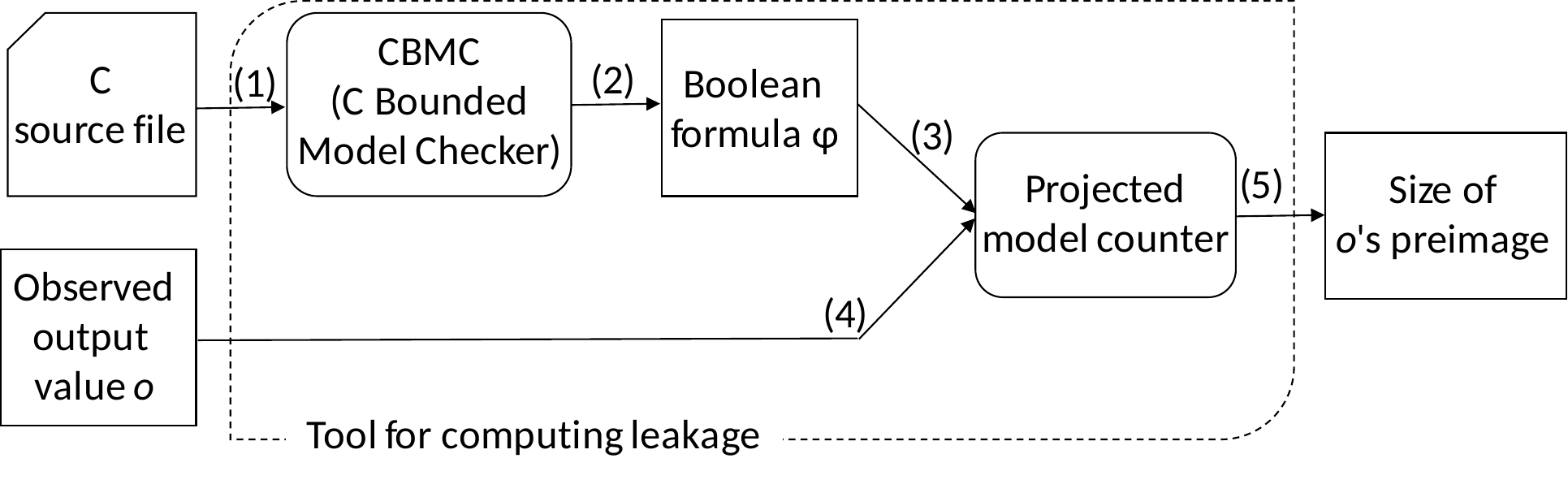,scale=0.4}\end{center}
	\caption{Reduction of computing dynamic leakage to model counting}
	\label{fig:3}
\end{figure}

In the remainder of Section 4, we assume input values are uniformly distributed for simplicity. Fig. \ref{fig:3} illustrates the calculation flow using model counting. The basic idea is similar to other existing QIF analysis tools based on model counting, 
namely, 
(1) feeding a target C program into CBMC \cite{CBMC}; 
(2) getting a Boolean formula $\varphi$ equivalent to the source program
in terms of constraints among variables in the program; 
(3) feeding $\varphi$ into a projected model counter that can count the models
with respect to projection on variables of interest; and 
(5) getting the result. 
The only difference of this framework from existing ones is (4), 
augmenting information about an observed output value $o$ into the Boolean formula $\varphi$ 
so that each model corresponds to an input value which produces $o$. 
The set of the obtained models is exactly the pre-image of $o$. 
\medskip\par\noindent
{\bf Pluggability} There are several parts in the framework above 
that can be flexibly changed to utilize the strength of different tools and/or approaches. 
Firstly, the {\it projected model counter} at (5) could be 
either a projected $\sharp$SAT solver (e.g., SharpCDCL) or a $\sharp$SMT solver (e.g., aZ3). 
Consequently, a formula at (3) could be
either a SAT constraints (e.g., a Boolean formula in DIMACS format) 
or a SMT constraints (e.g., a formula in SMTLIB format) generated by CBMC. 
Moreover, this framework can be extended to different programming languages other than C, such as Java, having JPF \cite{VHBPL03} and KEY \cite{DHS05} as two well-known counterparts of CBMC. 
In the next section, we are showing experimental results in which 
we tried several set-ups of tools in this framework to observe the differences.

\setcounter{section}{4}
\section{Experiments}
We conducted some experiments to investigate the flexibility of the framework to reduce computing dynamic leakage to model counting introduced in the previous section, as well as the scalability of this method. For the simplicity to achieve this purpose, we restricted to calculate dynamic leakage for {\it deterministic programs with uniformly distributed input}. Toward the analysis in more general cases and possibilities on performance improvement, we give some discussion and leave it as one of future work.

\subsection{Overview}
All experiments were done in a same PC with the following specification: core i7-6500U, CPU@2.5GHz x 4, 8GB RAM, Ubuntu 18.04 64 bits. We set one hour as time-out and interrupted execution whenever the running time exceeds this duration. The model counters we used are described below.
\begin{itemize}
	\item {\it aZ3}: a $\sharp$SMT solver developed by Phan et al. \cite{PM15}, which is built on top of the state-of-the-art SMT solver Z3. We used an improved version of aZ3 which is developed by Nakashima et al. \cite{NCHSS16}. It allows specifying variables of interest, which is equivalent to projection in SAT-based model counter. 
	\item {\it SharpCDCL}: a $\sharp$SAT solver with capability of projected counting based on Conflict-Driven Clause Learning (CDCL) \cite{SharpCDCL}. 
	The tool finds a new projected model and then adds a clause blocking to find the same model again. It enumerates all projected models by repeating that. 
	\item {\it DSharp-p}: another $\sharp$SAT solver based on d-DNNF format \cite{MMBH12}. 
	The tool first translates a given formula into d-DNNF format. It is known that, once given a d-DNNF format of constraints, it takes only linear time to the size of the formula to count models of those constraints. 
	We used an extended version with the capability of projected counting which is added by Klebanov et al. \cite{KMM13,DSharp-p}.
	\item {\it GPMC}: a projected model counter built on top of the SAT solver glucose \cite{Glucose}, in which component analysis and 
	caching used in the model counter SharpSAT are implemented \cite{GPMC}. 
\end{itemize}
The benchmarks are taken from previous researches about QIF analysis with most of them are taken from benchmarks of aZ3 \cite{PM15}, except {\it bin_search32.c} which is taken from \cite{MS11}. The difference between the ordinary QIF analysis and dynamic leakage quantification is that the former is not interested in an observed output value, but the latter is. Therefore, for the purpose of these experiments, we augmented the original benchmarks with additional information about concrete values of public output (public input also if there is some). Because we assume deterministic programs with uniformly distributed input, $\QIFone(o) = \QIFtwo(o) = \log \frac{|{\cal S}|}{|\pre_P(o)|}$ by Theorem \ref{th:det-quant}. Hence, without loss of precision in comparison, we consider counting $\pre_P(o)$ as the final goal of these experiments.
\subsection{Results}
Table \ref{tbl:exper1} shows execution time of model counting based on the four different model counters, in which {\it t/o} indicates that the experiment was interrupted because of time-out and - means the counter gave a wrong answer (i.e., only DSharp-p miscounted for UNSAT cases, probably because the tool does assume input formula to be satisfiable). By eliminating parsing time from the comparison, we measured only time needed to count models. 

According to the experimental results, aZ3-based model counting did not win the fastest for any benchmark, and moreover its execution time is always at least ten times slower than the best. On the other hand, DSharp-p seems to take much time to translate formulas into d-DNNF format for {\it dining6/50.c} and {\it grade.c}, and gave wrong answers for {\it mix\_duplicate.c} and {\it sanity\_check.c}, 
the number of models of which are 0, i.e., unsatisfiable. By and large, aZ3 and DSharp-p can hardly take advantage to the other tools, SharpCDCL and GPMC, in dynamic leakage quantification. 
Though SharpCDCL won 8 out of 14 benchmarks, the difference between the tools in those cases are not significant, yet the execution times are too short that it can be fluctuated by insignificant parameters. Therefore, it is better to look at long run benchmarks, {\it grade.c} and {\it masked\_copy.c}. In both cases, GPMC won by 9.7 times and 287.9 times respectively. The execution times as well as the difference in those two cases are significant. We also noticed that, those two cases have 65,536 and 65 models, which are the two biggest counts among the benchmarks. The more the number of the models is, the bigger is the difference between execution times of GPMC and SharpCDCL. Hence, we can empirically conclude that GPMC-based works much better than SharpCDCL-based in cases the number of models is large, while not so worse in other cases.

By implementing the prototype, we reaffirmed the possibility of automatically computing $\QIFone$ and $\QIFtwo$. Speaking of scalability, despite of small LOC (Lines of Code), there is still the case of {\it grade.c} (48 lines) for which all settings take longer than one minute, a very long time from the viewpoint of runtime analysis, to count models. 
There are several directions to improve the current performance which we leave as one of future work. First, because dynamic leakage should be calculated repeatedly for different observed outputs but a same program, we can leverage such an advantage of d-DNNF that while transforming to a d-DNNF format takes time, the model counting can be done in linear time once a d-DNNF format is obtained. That is, we generate merely once in advance a d-DNNF format of the constraints representing the program under analysis, then each time an observed output value is given, we make only small modification and count models in linear time to the size of the constraints. The difficulty of this direction lies in how to augment the information of observed output to the generated d-DNNF without breaking its d-DNNF structure. Another direction is to loose the required precision to accept approximate count. This could be done by counting on existing approximate model counters. 
\begin{table*}[!t]
	\caption{Counting result and execution time (ms) of different settings}
	\begin{center}
		\begin{tabular}{|c|c| c c c c|}
			\hline {\bf Benchmark} & {\bf Count} & {\bf aZ3} & {\bf SharpCDCL} & {\bf DSharp-p} & {\bf GPMC} \\
			\hline bin_search32 & 1 & 781 & 37 & 52 & {\bf 9} \\
			\hline crc8 & 32 & 303 & {\bf 11} & 31 & 36 \\
			\hline crc32 & 8 & 294 & {\bf 8} & 32 & 32 \\
			\hline dining6 & 6 & 1,305 & {\bf 44} & {\it t/o} & 49 \\
			\hline dining50 & 50 & {\it t/o} & 199 & {\it t/o} & {\bf 193} \\
			\hline electronic_purse & 5 & 525 & {\bf 137} & 9,909 & 223 \\
			\hline grade & 65 & 2,705,934 & 910,655 & {\it t/o} & {\bf 93,445} \\
			\hline implicit_flow & 1 & 253 & {\bf 15} & 31 & 33 \\
			\hline masked_copy & 65,536 & {\it t/o} & 9,214 & {\bf 30} & 32 \\
			\hline mix_duplicate & 0 & 241 & 12 & - & {\bf 4} \\
			\hline population_count & 32 & 477 & {\bf 19} & 37 & 34 \\
			\hline sanity_check & 0 & 247 & 13 & - & {\bf 8} \\
			\hline sum_query & 3 & 310 & {\bf 20} & 31 & 35 \\
			\hline ten_random_outputs & 1 & 249 & {\bf 18} & 32 & 34 \\
			\hline
		\end{tabular}
	\end{center}
	\label{tbl:exper1}
\end{table*}
\newline
\subsection{Toward general cases}
In order to calculate $\QIFone$ and $\QIFtwo$ for a probabilistic program with a non-uniform input distribution, 
we must identify projected models of the Boolean formula, rather than the number of the models,  
to obtain the probabilities determined by them in general. 
GPMC, specialized for model counting, does not compute the whole part of each model explicitly. 
Hence, GPMC is not appropriate for a calculation of the probability depending on the concrete models. 
On the other hand, sharpCDCL basically enumerates all projected models, and thus 
we think we can extend it as follows to compute $\QIFone$ and $\QIFtwo$ in general cases.

To calculate $\QIFone$ for a probabilistic program with a non-uniform input distribution, 
we can replace each probabilistic choice in a give program with a non-deterministic 
choice as stated in Section 4, and then enumerate projected models with respect 
to the input variables, summing up the probabilities of the corresponding input values. 

As for $\QIFtwo$, we have to calculate not only the probabilities of possible input values 
but also those of possible execution paths reachable to the observed output. 
To achieve this,  in addition to the replacement of probabilistic choices with non-deterministic choices, 
we may insert variables to remember which branch is chosen at each of the non-deterministic choices. 
Then, given a projected model of the Boolean formula generated from the modified source code 
with respect to the input variables and the additional choice variables,
we can get to know a possible input value and an execution path from the projected model. 
For a possible input value $s$,  $p(o|s)$ is the sum of the probabilities of all possible execution paths 
from $s$ to the observed $o$. 

\section{Conclusion}
In this paper, we summarize three requirements as criteria for reasonable dynamic leakage definitions to follow. Also we defined two novel ones both of which satisfy all the criteria and have understandable explanations of the background perspectives. Besides giving proof of some of their characteristics, we gave results on the hardness of computing dynamic leakage under those definitions for three classes of Boolean programs, including loop-free, while and recursive. Despite of the hardness, we introduced a framework to reduce the problems to model counting, which gets much attention from researchers from various fields of interest. Based on that framework, we implemented a prototype and conducted some experiments to verify flexibility and scalability of the framework. Lastly, we gave some discussion on how to improve the performance and the whole picture of computing dynamic leakage in general cases.
\newline
Beyond this paper, we leave the following as future work: (1) utilizing the strength of d-DNNF format to improve calculation performance, (2) approaching those problems in terms of approximated calculation and (3) tackling the problems under more general assumptions. 
%
%

\begin{thebibliography}{8}
	\bibitem{AEY01}
	R.\ Alur, K.\ Etessami, M.\ Yannakakis, 
	\textit{Analysis of recursive state machines}, 
	13th Intenational Conference on Computer-Aided Verification (CAV), 2001, 304--313. 
	
	\bibitem{ACM16}
	M.\ S.\ Alvim, K.\ Chatzikokolakis, A.\ McIver, 
	\textit{Axioms for information leakage}, 
	29th Computer Security Foundations Symposium (CSF), 2016, 77--92.
	
	\bibitem{ACMS15}
	R.\ A.\ Aziz, G.\ Chu, C.\ Muise, P.\ Stuckey,
	\textit{\#$\exists$SAT: projection model counting},
	18th International Conference on Theory and Applications of Satisfiability Testing (SAT), 2015, 121--137.
	
	\bibitem{BBJ16}
	F.\ Besson, N.\ Bielova, T.\ Jensen,
	\textit{Hybrid Monitoring of Attacker Knowledge},
	29th Computer Security Foundations Symposium (CSF), 2016, 225--238.
	
	\bibitem{Bi16}
	N.\ Bielova,
	\textit{Dynamic leakage - a need for a new quantitative information flow measure},
	ACM Workshop on Programming Languages and Analysis for Security (PLAS), 2016, 83--88.
	
	\bibitem{BEHLMQ18}
	F.\ Biondi, M.\ A.\ Enescu, A.\ Heuser, A.\ Legay, K.\ S.\ Meel, J.\ Quilbeuf,
	\textit{Scalable approximation of quantitative information flow in programs},
	Verification, Model Checking, and Abstract Interpretation (VMCAI), 2018, 71--93.
	
	\bibitem{BKLT17}
	F.\ Biondi, Y.\ Kawamoto, A.\ Legay, L.\ M.\ Traonouez,
	\textit{HyLeak: hybrid analysis tool for information leakage},
	Automated Technology for Verification and Analysis (ATVA), 2017, 156--163.
	
	\bibitem{BEM97}
	A.\ Bouajjani, J.\ Esparza, O.\ Maler,
	\textit{Reachability analysis of pushdown automata: application to model-checking},
	8th International Conference on Concurrency Theory (CONCUR), 1997, 135--150.
	
	\bibitem{CU12}
	R.\ Chadha and M.\ Ummels,
	\textit{The complexity of quantitative information flow in recursive programs},
	Research Report LSV-2012-15, Laboratoire Sp\'{e}cification \& V\'{e}rification, \'{E}cole Normale Sup\'{e}rieure de Cachan, 2012. 
	
	\bibitem{CKN14}
	T.\ Chothia, Y.\ Kawamoto, C.\ Novakovic,
	\textit{LeakWatch: estimating information leakage from Java programs},
	19th European Symposium on Research in Computer Security
	(ESORICS), 2014, 219--236.
	
	\bibitem{CMS09}
	M.\ R.\ Clarkson, A.\ C.\ Myers and F.\ B.\ Schneider,
	\textit{Quantifying information flow with beliefs},
	18th Computer Security Foundations Symposium (CSF), 2009, 655--701.
	
	\bibitem{DHS05}
	A.\ Darvas, R.\ H\"{a}hnle, D.\ Sands,
	\textit{A theorem proving approach to analysis of secure information flow},
	Security in Pervasive Computing (SPC), 2005, 193--209.
	
	\bibitem{EY05}
	K.\ Etessami, M.\ Yannakakis,
	\textit{Recursive Markov chains, stochastic grammars, and monotone systems of nonlinear equations},
	Journal of ACM (JACM), Vol. 56, Issue 1, Jan 2009.
	
	\bibitem{GM82}
	J.\ A.\ Goguen, J.\ Meseguer,
	\textit{Security policies and security models},
	IEEE Symposium on Security and Privacy (S$\&$P), 1982, 11--20.
	
	\bibitem{KMM13}
	V.\ Klebanov, N.\ Manthey, C.\ Muise,
	\textit{SAT-based analysis and quantification of information flow in programs},
	Quantitative Evaluation of Systems (QEST), 2013, 177-192.
	
	\bibitem{KR10}
	B.\ K\"{o}pf, A.\ Rybalchenko,
	\textit{Approximation and randomization for quantitative information flow analysis},
	23rd Computer Security Foundations Symposium (CSF), 2010, 3--14.
	
	\bibitem{ME08}
	S.\ McCamant, M.\ D.\ Ernst,
	\textit{Quantitative information flow as network flow capacity},
	ACM SIGPLAN Conference on Programming Language Design and Implementation (PLDI), 2008, 193--205.
	
	\bibitem{MS11}
	Z.\ Meng, G.\ Smith,
	\textit{Calculating bounds on information leakage using two-bit patterns},
	6th Workshop on Programming Languages and Analysis for Security (PLAS), 2011, 1--12.
	
	\bibitem{MU05}
	M.\ Mitzenmacher, E.\ Upfal,
	\textit{Probability and computing: randomized algorithms and probabilistic analysis}, 
	Cambridge, 2005, 167--173.
	
	\bibitem{MMBH12}
	C.\ Muise, S.\ A.\ McIlraith, J.\ C.\ Beck, E.\ Hsu,
	\textit{DSHARP: fast d-DNNF compilation with sharpSAT},
	Advances in Artificial Intelligence (AI), 2012, 356--361.
	
	\bibitem{NCHSS16}
	S.\ Nakashima, B.\ T.\ Chu, K.\ Hashimoto, M.\ Sakai, H.\ Seki,
	\textit{Efficiency improvement in $\sharp$SMT-based quantitative information flow analysis},
	IEICE Technical Report, SS2016-26, Vol. 116, No. 277, 2016, 49--54.
	
	\bibitem{PM15}
	Q.\ S.\ Phan, P.\ Malacaria,
	\textit{All-solution satisfiability modulo theories: applications, algorithms and benchmarks},
	10th International Conference on Availability, Reliability and Security (ARES), 2015, 100--109.
	
	\bibitem{Sm09}
	G.\ Smith,
	\textit{On the foundations of quantitative information flow},
	12th International Conference on Foundations of Software Science and Computational Structures (FOSSACS), 2009, 288--302.
	
	\bibitem{SHS17}
	R.\ Suzuki, K.\ Hashimoto, M.\ Sakai,
	\textit{Improvement of projected model-counting solver with component decomposition using SAT solving in components},
	JSAI Technical Report, SIG-FPAI-506-07, 2017, 31--36 (in Japanese).
	
	\bibitem{VEBAH16}
	C.\ G.\ Val, M.\ A.\ Enescu, S.\ Bayless, W.\ Aiello, A.\ J.\ Hu,
	\textit{Precisely measuring quantitative information flow: 10k lines of code and beyond},
	IEEE European Symposium on Security and Privacy (EuroS$\&$P), 2016, 31--46.
	
	\bibitem{VHBPL03}
	W.\ Visser, K.\ Havelund, G.\ Brat, S.\ J.\ Park, F.\ Lerda,
	\textit{Model checking programs},
	Automated Software Engineering (ASE), Vol. 10, Issue 2, 2003, 203--232.
	
	\bibitem{YT10}
	H.\ Yasuoka, T.\ Terauchi,
	\textit{Quantitative information flow - verification hardness and possibilities},
	23rd Computer Security Foundations Symposium (CSF), 2010, 15--27.
	
	\bibitem{YT11}
	H.\ Yasuoka, T.\ Terauchi,
	\textit{On bounding problems of quantitative information flow},
	Journal of Computer Security (JCS), Vol. 19, 2011 November, 1029--1082.
	
	
	\bibitem{CBMC}
	C Bounded Model Checker,
	\url{https://www.cprover.org/cbmc}.
	
	\bibitem{DSharp-p}
	DSharp-p,
	\url{https://formal.iti.kit.edu/~klebanov/software/}
	
	\bibitem{Glucose}
	Glucose SAT Solver,
	\url{https://www.labri.fr/perso/lsimon/glucose}.
	
	\bibitem{GPMC}
	GPMC,
	\url{https://www.trs.css.i.nagoya-u.ac.jp/~k-hasimt/tools/gpmc.html}.
	
	\bibitem{SharpCDCL}
	SharpCDCL,
	\url{http://tools.computational-logic.org/content/sharpCDCL.php}.
\end{thebibliography}
%

\end{document}